# Effects of the Charge-Dipole Interaction on the Coagulation of Fractal Aggregates

Lorin Swint Matthews and T. W. Hyde, *Member, IEEE*

*Abstract*—A numerical model with broad applications to complex (dusty) plasmas is presented. The self-consistent N-body code allows simulation of the coagulation of fractal aggregates, including the charge-dipole interaction of the clusters due to the spatial arrangement of charge on the aggregate. It is shown that not only does a population of oppositely charged particles increase the coagulation rate, the inclusion of the charge-dipole interaction of the aggregates as well as the electric dipole potential of the dust ensemble decreases the gelation time by a factor of up to twenty. It is further shown that these interactions can also stimulate the onset of gelation, or "runaway growth," even in a population of particles charged to a monopotential where previously it was believed that like-charged grains would inhibit coagulation. Gelation is observed to occur due to the formation of high-mass aggregates with fractal dimensions greater than two which act as seeds for runaway growth.

## I. Introduction

THE coagulation of micrometer sized particles in a complex (dusty) plasma is a fundamental process that has become an increasingly important area of study in the last ten years, not only in the context of astrophysical systems [1], but also in many parts of chemistry, physics, and colloidal systems such as aerosols. At the same time, as feature size continues to decrease, particle coagulation has also become germane to contamination problems in the fabrication of semiconductor wafers through plasma processing [2].

Interest in coagulation in astrophysical systems has been spurred by recent discoveries of extrasolar planets, renewing attention to the physics of planet formation. The most widely accepted theory of the origin of planets is that they formed through mutual collisions of planetesimals, bodies ranging in size from 1 to 10 km [3]. The planetesimals in turn are thought to have formed from the dominant constituent material of circumstellar disks; gas and dust [4]. With the advent of the Hubble Space Telescope, direct observations of the gas/disk systems around T Tauri stars have dramatically increased interest in the coagulation of dust within astrophysical environments [5], [6].

As the gas in a protoplanetary disk cools, coagulation of dust particles begins as inelastic, adhesive collisions occur. The primary factors that affect the coagulation rate are the relative velocity between grains, collisional cross-section, and sticking probability. Several experimental and numerical studies have made it evident that coagulation results in the formation of fluffy fractal aggregates [7] – [9]. Fractal aggregates exhibit stronger gas-grain coupling, suppressing the relative velocities between aggregates [10], and have greater collisional cross sections due to their open nature. This has been shown to initially increase the coagulation rate [11], [12] and can lead to "runaway cluster growth" where a single aggregate rapidly accumulates a large fraction of the mass in the system [13], [14]. Thus the physical geometry of the dust grains is an important factor in properly modeling dust coagulation.

In the primordial solar nebula, the dust is imbedded in a plasma and can become charged. For low dust densities, the greater thermal speeds of the electrons result in a higher flux of electrons than ions and consequently the dust grains become negatively charged. For isolated grains in a hydrogen plasma, the equilibrium surface potential is on the order of -2.51 kT/e [15], where $T$ is the temperature of the plasma. The resulting repulsive grain potential necessarily leads to reduced coagulation rates, though it is still possible for coagulation to occur as long as the dust grains have relative velocities large enough to overcome the Coulomb barrier, yet smaller than the critical sticking velocity [16].

The above can be moderated by temperature fluctuations and differences in the secondary electron yield, which can lead to the formation of dust populations with oppositely charged grains depending on the size of the grains [17] – [21]. This in turn can lead to enhanced coagulation rates. Evidence of rapid charge-induced coagulation has also been observed in a neutral gas environment, where particles or clusters can become charged through collisions [14].

A variety of works have studied the effects of various parameters on the coagulation of dust particles. Many of these employ statistical techniques to solve the coagulation equation [11], [14], [22]. However, the complexity of the problem has shown that it is advantageous to use a self-consistent N-body approach to fully model and understand the processes involved [12],[23]. The coagulation of μm-sized particles has also been the subject of several recent experimental studies (e.

Manuscript received July 15, 2003. This work was supported in part by the National Science Foundation grant PHY-0097386.

L. S. Matthews and T. W. Hyde are with the Center for Astrophysics, Space Physics, and Engineering Research, Baylor University, Waco, TX 76798-7310 USA (phone: 254-710-3763; fax: 254-710-7309; e-mail: lorin_matthews@baylor.edu).

g. [7], [24], [25]). Most of these have looked at the coagulation of a monodisperse size population, given that coagulation due to Brownian motion tends to form cluster of similar size [23]. Only a few groups have examined the effects of particle charge on coagulation [11], [14], [17], [22].

In this study, we employ a self-consistent N-body code to model the effects of grain charge on coagulation. By modeling a dust population with a given size distribution, the effects of a population with oppositely charged grains can be compared to those with grains that are neutral or charged to a mono-potential. The code also allows the three-dimensional geometry of the fractal aggregates to be tracked, enabling the investigation of the effects of the distribution of the charge on fractal aggregates, including the charge dipole moment and it influence on the force calculation.

## II. NUMERICAL MODEL

### A. Overview

The numerical model employed in this study is based on the box_tree code developed by Richardson [12], [26], [27] to model the dynamics of a large number of particles in a protoplanetary disk or ring system interacting through gravitational forces. The box_tree code is a hybrid of two computer algorithms, a box code and a tree code. The box code, first used to study dynamic ring systems [28], allows distant regions of a system to be represented by copies of the simulated region. The box code, in providing the external potentials acting on the grains, specifies a coordinate system, the linearized equations of motion, and a prescription for handling boundary conditions. The tree code [29] provides a method for a fast calculation of the interparticle forces by means of a multipole expansion. The interparticle forces are then included as a perturbation to the equations of motion. Using this code, a full treatment of rigid body dynamics, including rotation, is possible allowing for both cluster trajectories and the orientation of fractal aggregates to be followed. Additional modifications were made to the code to allow for charges on the grains, electrostatic Debye shielding, external magnetic fields and the charge dipole interaction of fractal aggregates. Thus a variety of astrophysical environments can be examined, from free space to rotating disk systems [30] – [33] as well as strongly coupled complex plasmas such as dust crystals and coulomb clusters [34] - [37].

### B. Multipole Expansion of the Tree

As part of the tree code, the box containing the particles in the system is divided into $2^n$ cells, where n is the number of dimensions of the box. Each cell is further divided into $2^n$ cells until each cell contains only a single particle. If a cell is sufficiently far away from the grain of interest, as defined by a critical opening angle [27], the multipole moments calculated using the particles in that cell are employed to calculate the self-gravitational and electrostatic forces on the grain. Otherwise the forces due to the grains within a cell are calculated directly without the use of a multipole expansion.

The multipole terms for the tree structure for the self-gravitational and electrostatic forces are expanded about a cell's center of mass. The gravitational dipole moment of the cell is zero due to this choice of origin; however the electrostatic dipole moment is non-zero for both oppositely charged and like-charged populations. Debye shielding of the grain potentials is incorporated into the code by scaling the electrostatic force, $F_e$, using a screening factor

$$F'_e = F_e \left(1 + r/\lambda_d \right) e^{-r/\lambda_d} \quad (1)$$

where $\lambda_d$ is the Debye length of the plasma and $r$ is the distance from a grain.

### C. Collision Resolution

As mentioned above, the outcome of a collision between grains depends strongly on the energies involved. Two colliding particles may bounce, stick, fragment, crush, melt, or vaporize. The physics of the collision process was developed by Chokshi et al. [16] in which they calculated the critical velocity for sticking, $v_{cr}$, depending on a grain's physical characteristics. They found that colliding particles must have a relative velocity less than $v_{cr}$ in order for sticking to occur.

For particles with radii less than $10^{-4}$ m, Brownian motion is the dominant source of collisions among particles [8]. Brownian motion also dominates the motion of fractal aggregates up to approximately 10 μm in size, or a few hundred 1 μm particles. Thus, thermal motion leads to a Maxwellian velocity distribution with a mean thermal velocity given by [11]

$$\langle v_{th} \rangle = \sqrt{8kT/\pi m} \quad (2)$$

where $T$ is the kinetic temperature, $k$ is the Boltzmann constant, and $m$ is the mass of the particle. Calculating the average relative velocity between the grains,

$$\langle v \rangle = \sqrt{8kT(m_1 + m_2)/\pi m_1 m_2} \,, \quad (3)$$

using temperatures and masses appropriate for nebular clouds or protoplanetary disks, it can be seen that the average relative velocity will always be less than the critical velocity for sticking [11]. Thus the energies required for restructuring are probably not available during the early stages of dust particle growth.

The box_tree code allows for two primary types of merging events. In a simple merge, two spherical particles collide and form a single spherical particle having a mass $m = m_1 + m_2$. Linear and angular momentum about the center of mass of the colliding particles are conserved. The second option allows the particles to form fractal aggregates in which the colliding particles stick together at the point of contact. Linear and angular momentum are conserved during the collision by calculating the moment of inertia of the resultant particle and then utilizing Euler's equations for rigid body rotation [12].

*D. Rearrangement of Charge on Fractal Aggregates*

For spherical particles, the charge (consisting of free electrons and ions) is evenly distributed over the surface with the potential of the charged grain that of a point charge located at the center of the particle. However, on fractal aggregates the charge is allowed to rearrange itself in such a way as to minimize the overall potential on the surface. Therefore it is unrealistic to treat the aggregate's potential as that of a point charge residing at the aggregate's center of mass. It would be quite difficult, as well as extremely time consuming, to explicitly calculate the exact potential of each fractal aggregate. Thus some approximation of the distribution of charge on the aggregate is in order.

Due to the mutual repulsion between charges on the grain surface, the majority of the charge on the particle should reside on the aggregate's extremities. Additionally, because of the "open" nature of the fractal, the central particles in the aggregate will also have exposed surface area, and some fraction of the charge could be expected to reside there as well. As a first approximation for dealing with this problem, box_tree assumes that each monomer in the aggregate has a charge

$$q = QD_i/D_T \qquad (4)$$

where Q is the total charge on the aggregate, $D_i$ is the distance of the i$^{th}$ monomer from the center of mass of the aggregate, and $D_T$ is given by

$$D_T = \sum D_i . \qquad (5)$$

Thus particles farther away from the center of mass have a larger charge than do particles near the center of the aggregate.

*E. Multipole Expansion of the Aggregate*

The overall potential of the aggregate is calculated via a multipole expansion about the center of mass, with this result then used to calculate the force on all other particles. The dipole-dipole interactions between two fractal aggregates are not calculated, rather the charge of the second aggregate is assumed to reside solely at the center of mass. It is also assumed that all of the charge resides at the aggregate's center of mass when calculating the multipole moments of the tree. The orders of the multipole expansion to be used in the force calculations are specified at run time.

## III. SIMULATIONS

*A. Initial Conditions*

As mentioned above, the coagulation of micrometer sized dust particles begins early in the formation of protostellar clouds and continues through the early stages of protoplanetary disk development. Since the scale of the box size within the numerical model is small when compared against the scale of a protostellar or protoplanetary system, dust grain coagulation can be modeled as taking place in free space in a non-rotating frame. Particles were given an initial uniform random distribution and velocity dispersion with 5000 particles having radii ranging in size from 1 – 6 μm examined. A particle number density n = $10^5$/cm$^3$ was assumed, which is comparable to the dust density at the center edge of a protostellar cloud and about an order of magnitude greater than the density at the outer edge [38]. Since the Debye length in protostellar clouds is much greater than the average interparticle distance, Debye shielding has little effect on the aggregation process and was omitted from this study. For illustrative purposes, the grains' potentials were chosen to correspond to those of Horanyi and Goertz [17] in which temperature fluctuation or transient heating events were assumed to introduce a charge separation in the dust population. Grains with radii less than 4 μm have an average surface potential of -1V, while those with radii greater than or equal to 4 μm have an average surface potential of +4V (model 1). The resulting coagulation rate was compared to that determined for populations of uncharged grains (model 2) and like-charged grains with surface potentials of -1V (model 3). For each of the three models, four runs with different random number seeds were performed and the results then averaged together. As a first approximation and in order to speed up the computation time, only the monopole moments of the tree were used. In the original three models, charge was not distributed over the fractal aggregate, but was assumed to reside at the aggregate's center of mass.

Four additional models were used to study the effects of including the electric dipole moments of the fractal aggregates as well as the tree structure. Models 4 and 5 consisted of oppositely charged grains as in model 1. Only the dipole moment of the aggregates was used in the force calculations of model 4, while model 5 included the dipole moments of the tree structure as well. Models 6 and 7 correspond to model 3 (like-charged grains) with the dipole moments of the aggregate and dipole moments of the aggregates and tree structure included in the force calculations, respectively. Due to the CPU time required for these simulations (ranging from 10 days to two months on a DEC Alpha EV67 PS 667) only one run was conducted for each of these models. A summary of the models is given in Table I.

TABLE I
SUMMARY OF TEST MODELS

| Model | CHARGE OF POPULATION | Multipole Moments Used in Force Calculation | |
|---|---|---|---|
| | | Fractal Aggregate | Tree Structure |
| 1 | Oppositely charged | Monopole | Monopole |
| 2 | Neutral | Monopole | Monopole |
| 3 | Like-charged | Monopole | Monopole |
| 4 | Oppositely charged | Dipole | Monopole |
| 5 | Oppositely charged | Dipole | Dipole |
| 6 | Like-charged | Dipole | Monopole |
| 7 | Like-charged | Dipole | Dipole |

As mentioned in Section II, grains must have relative velocities large enough to overcome the Coulomb barrier in order for coagulation to occur. For 1 μm grains charged to a -1V potential, this minimum relative velocity is ~16 cm/s while for a 6 μm grain charged to a 4V potential this velocity is 2.7 cm/s. Although this is somewhat greater than the thermal velocities expected due to Brownian motion, turbulence in a protoplanetary disk can easily lead to such velocities [39]. As a result, the grains in this model were given initial velocities on the order of 10 cm/s. Laboratory experiments on the aggregation of μm-sized solid grains and dust agglomerates have shown that the sticking efficiency is always unity for moderate collision velocities v ≤ 1 m/s and that monomers merge to form fractal aggregates [24]. Therefore, the collisions in the simulation were required to produce fractal aggregates with no restructuring or restitution.

### B. Size Distribution of the Population

A defining characteristic of collisional growth is the evolution of the mean aggregate mass of the population over time. The mean aggregate mass can be expected to grow as $m \propto t^2$ if the collisional cross-section is proportional to the aggregate mass [23] and as $m \propto \exp(t)$ which results when the collision cross section scales as $m^{2/D_f}$, where $D_f$ is the fractal dimension, described in the next section [40].

The evolution of the mass of the aggregates (normalized to the mean monomer mass, $m_o$) versus time is plotted for models 1-3 in Fig. 1. Increased coagulation can be seen in model 1 (oppositely charged grains) as compared to that of model 2 (neutral grains). At t = 5.0 s, model 1 had 100% of the mass aggregated with one cluster accounting for an average of 31±14% of the total mass. This is in comparison to model 2, which had 98% of the particles incorporated into an aggregate structure with the largest aggregate accounting for 3.4±0.6% of the total mass and model 3, which also had 98% of the of the particles incorporated and a mean largest mass of 1.1±0.2% at the same elapsed simulation time. As expected, a population of like-charged grains retards the coagulation process.

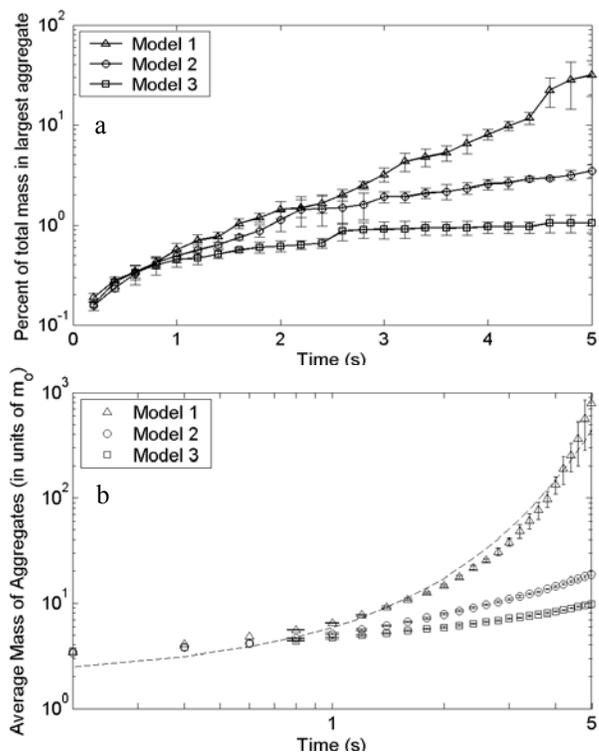

Fig. 1. Temporal evolution of the aggregate size in models 1-3. a) Evolution of the aggregate with the largest mass. b) Evolution of the average mass of all aggregates. The dashed line in is an exponential fit to the data for model 1 indicating exponential growth.

The evolution of the differential mass distribution functions is shown for models 1-3 in Fig. 2. The mass fraction per logarithmic interval, $N_i = \sum_j m_{ij}/m_{tot}$, is plotted where $i$ indicates the $i$th logarithmic bin and $j$ is the index of each aggregate populating the bin. The data from the individual runs was binned and averaged for each model. The distributions are given for times $t_1 = 0.2$ s, $t_2 = 1.0$ s, $t_3 = 2.0$ s, $t_4 = 3.0$ s, and $t_5 = 4.0$ s. The evolution of the mass distributions for both models 2 and 3 compare qualitatively with those seen in experimental studies for uncharged grains [24]. In contrast, the mass distribution functions for model 1 rapidly shift to larger masses and develop the high mass tail indicative of the onset of gelation, or "runaway growth". This corresponds to the formation of a few aggregates with large masses that have incorporated most or all of the small-mass aggregates.

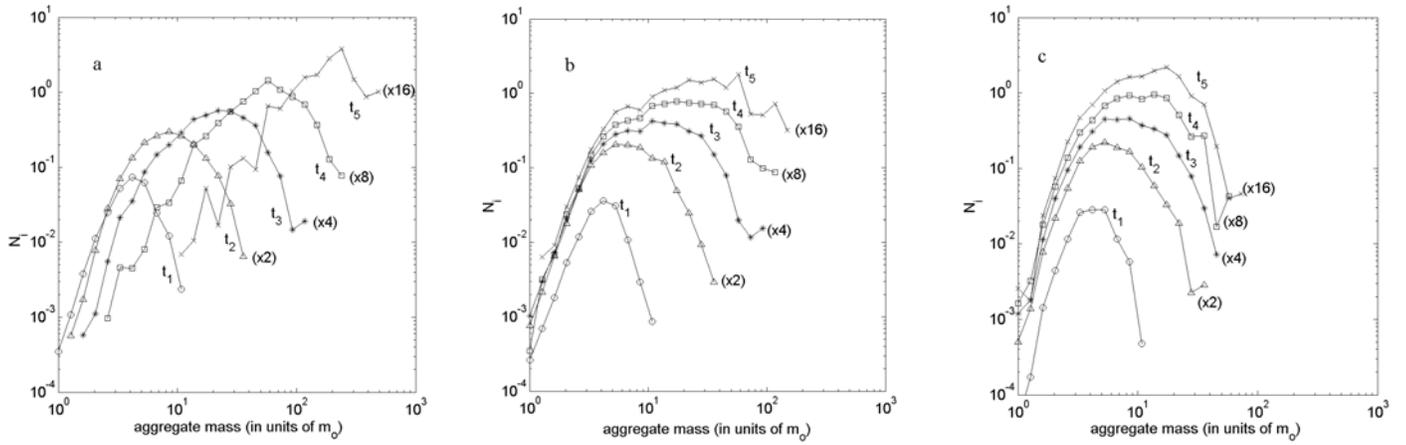

Fig. 2. Evolution of differential mass distribution function at times $t_1 = 0.2$ s, $t_2 = 1.0$ s, $t_3 = 2.0$ s, $t_4 = 3.0$ s, and $t_5 = 4.0$ s for a) model 1, b) model 2, and c) model 3. The population of oppositely charged grains (model 1) rapidly shifts to higher masses with a high mass tail. Subsequent curves have been shifted by a factor of 2 for clarity.

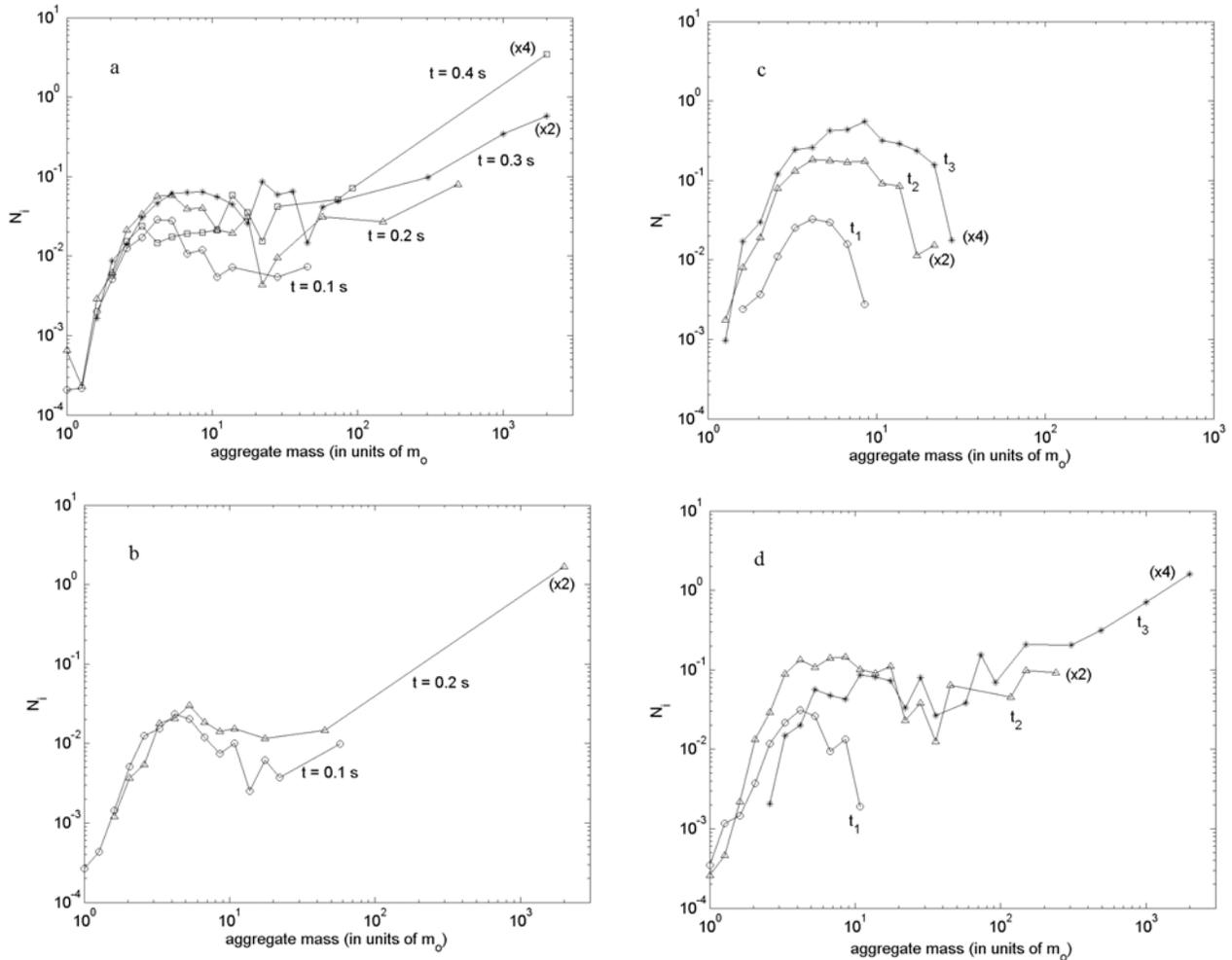

Fig. 3. Evolution of mass distributions for a) model 4, b) model 5, c) model 6 and d) model 7. Data is shown for times $t_1 = 0.2$ s, $t_2 = 1.0$ s, $t_3 = 2.0$ s in c) and d). Some curves have been shifted by factors of two for clarity.

The mass fraction distributions for models 4-7 are shown in Fig. 3. As can be seen, the coagulation for oppositely charged grains takes place on a much shorter time scale with the onset of gelation occurring in less than a second. When the dipole moment of the fractal aggregates is considered, the gelation time is decreased by a factor of ten (compare Fig. 3a to Fig. 2a) The inclusion of the dipole moment of the tree structure further decreases this time by a factor of two (Fig. 3b). Including the dipole moment of the fractal aggregates alone has very little effect on the coagulation rate (compare Figs. 2c and 3c). However, for like-charged grains, including the dipole moment of the tree structure as well (Fig. 3d) stimulates the production of a high-mass tail with subsequent runaway growth, as seen in the populations with oppositely charged grains. The gelation time for like-charged grains is about ten times that for oppositely charged grains.

*C. Structure of Aggregates*

The growth process is strongly dependent on the fractal dimension of the aggregate structures [40]. The fractal dimension, $D_f$, is related to the size and number of constituent particles by

$$N(r) \propto r^{D_f} \tag{6}$$

where $N(r)$ is the number of grains within a volume of radius r [41]. Depending on the accretion method, the fractal dimension can range from $1.3 < D_f < 3$. Models of ballistic particle-cluster accretion (BPCA) produce dense clusters with $D_f \approx 3$ [42], while ballistic cluster-cluster accretion (BCCA) predicts fluffier aggregates with $1.7 < D_f < 2.1$ [43]. Laboratory experiments have shown aggregate growth follows BCCA [7], though microgravity experiments have produced aggregates with very low fractal dimension of 1.3 [9], [25]. The formation of a large compact aggregate can act as a seed for "runaway growth", a well-known phenomenon in aggregation theory [13], [44].

The fractal dimension of the aggregates formed in this simulation is calculated from the slope of a log-log plot of radius measured from the center of mass versus the enclosed particle number [12]. The values of $D_f$ obtained by this method must be considered as estimates and are less reliable for small aggregates, as the aggregates are finite structures and $D_f$ is only strictly defined for objects extending to infinity [45].

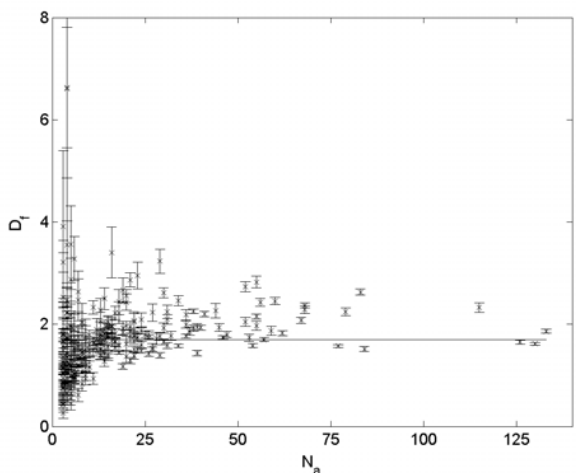

Fig. 4. Distribution of fractal dimension for all aggregates at time t = 5.0s for a single run of model 2. The error bars show the uncertainties in the calculated values.

A sample plot of $D_f$ versus $N_A$, the number of monomers in an aggregate, is shown in Fig 4. The data plotted is for all aggregates consisting of three or more monomers and is taken from a single run of model 2 (neutral grains) at time t = 5.0 s. The uncertainties in the calculation of $D_f$ are shown by error bars, which are quite large for small aggregates. The values of $D_f$ for large aggregates converge near two, while the mean value of $D_f$ for the population is $1.7 \pm 0.7$.

A comparison of the results for models 1-3 is shown in Fig. 5. In Fig. 5a-c, the probability density estimate of the fractal dimension is shown for aggregates with more than 10 monomers at time t = 4.0 s. The data for the four runs in each of the models was binned according to $D_f$ and then averaged, with error bars denoting the standard deviation of the mean. Model 1, which has undergone substantial coagulation, has a very broad distribution with 40% of aggregates having a fractal dimension greater than two. Models 2 and 3 have narrower distributions which peak at lower fractal dimensions. Only 20% and 14% of the aggregates in models 2 and 3, respectively, have aggregates with $D_f > 2$.

The mean size of the aggregates contributing to the bins is shown in Fig 6. The mean maximum and minimum cluster mass contributing to each bin is also indicated. These broad distributions are typical of those seen in both numerical [23] and experimental studies [24]. However, model 1 also shows several large-mass clusters with high fractal dimensions. (Note the difference in scale for the three graphs.) These dense clusters act as seeds for runaway growth.

A comparison of the fractal dimension of the aggregates for models 4-7 is shown in Fig 7. It is readily apparent that a significant fraction of the aggregate population in models 4, 5, and 7 have $D_f > 2$, while model 6 has a probability density that is centered around a lower fractal dimension (Fig 7a). Furthermore, models 4, 5, and 7 all have high mass clusters with $D_f > 2$, which act as seeds for runaway growth (Fig. 7b). While this is not unduly surprising for populations of oppositely charged grains, which are expected to exhibit enhanced coagulation, is it surprising that even populations of like-charged grains (e. g. model 7) can exhibit this behavior. It is clear that the electric dipole moment of the dust ensemble as well as that for fractal aggregates is responsible for this behavior. This has important implications for coagulation in a protoplanetary disk, as the demonstration that larger bodies can form from the constituent dust is yet an open issue [1]. Rapid coagulation due to runaway growth of charged grains, whether like-charged or oppositely charged, may prove to be a necessary ingredient for this process.

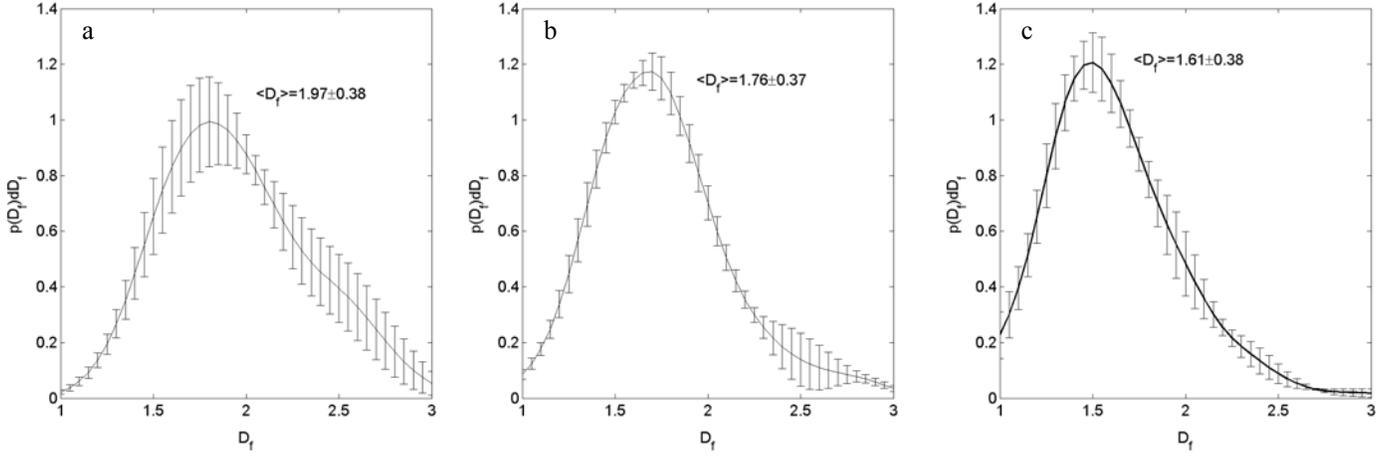

Fig. 5. Normalized probability density estimate of fractal dimension for aggregates consisting of more than ten monomers. Data taken at time t = 4.0 s for a) model 1 – oppositely charged grains, b) model 2 – neutral grains, and c) model 3 – like-charged grains.

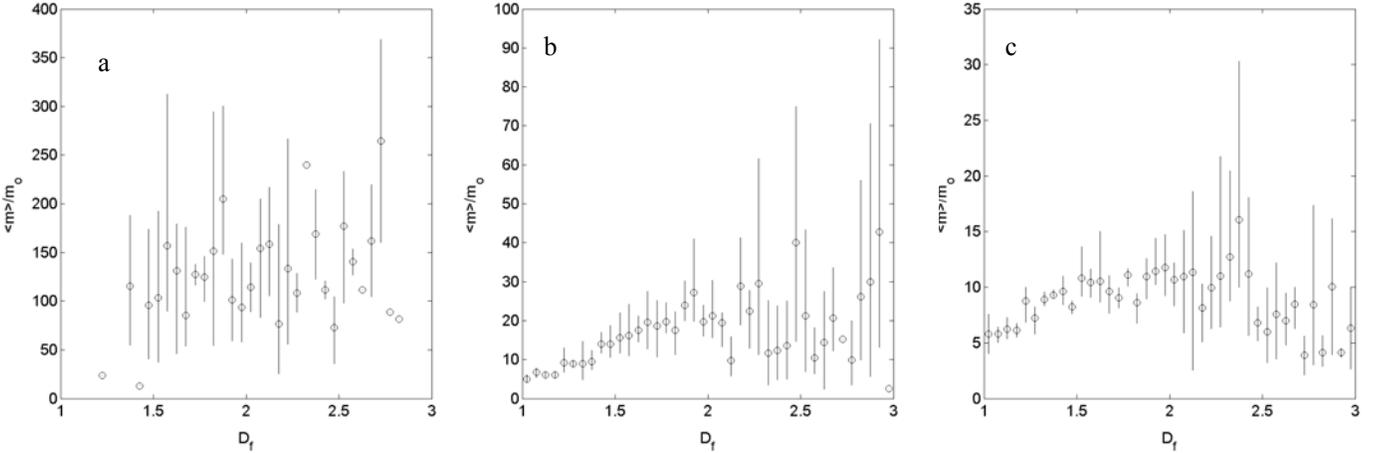

Fig. 6. Mean cluster mass (open circles), in bins for fractal dimension at t = 4.0s for a) 1, b) model 2, and c) model 3. The size variation of the clusters contributing to each bin is also indicated.

## IV. CONCLUSION

A new numerical model with broad applications to complex (dusty) plasma physics has been presented. This self-consistent fractal-aggregate model allows accurate numerical simulations of coagulation, while including the calculation of electrostatic forces between charged grains, the inclusion of screened potentials through Debye shielding, and the rearrangement of charge on fractal aggregates. The model produces a full treatment of rigid body dynamics, including rotation, enabling the examination of both cluster trajectories and the orientation of fractal aggregates. The coordinate system, particle size, particle charge, dust density, Debye length of the plasma, external potentials, and interparticle forces can all be specified by the user. The algorithm is robust, allowing application to a variety of interesting problems such as coagulation within rotating (protoplanetary/protostellar disks) and non-rotating frames (protostellar clouds), and the effects of charge and fractal agglomeration on coagulation.

It has been shown that populations of grains that become oppositely charged due to a dust size distribution have significantly enhanced coagulation rates, when compared to similar populations of like-charged or neutral grains. This is in agreement with previous statistical studies [17]. The mass distribution among fractal dimension, distribution of fractal dimensions, and temporal evolution of mass distribution functions also have been shown to agree with those found in other self-consistent N-body codes [23] and experimental studies [24].

Horányi and Goertz [17] assumed that oppositely charged grains could only occur during large fluctuations in plasma temperatures caused by transient heating events such as lightning, magnetic reconnection, or shocks, and thus enhanced coagulation rates would be intermittent at best. The fact that oppositely charged grains can also occur in relatively cool plasmas as a result of the size distribution of the dust in the plasma implies that enhanced coagulation due to oppositely charged grains may play a large role in the evolution of the protoplanetary disk.

Of even greater significance is the decrease in gelation time due to the charge-dipole interaction of the fractal aggregates as well as the electric dipole moment of the dust ensemble (as modeled by the tree structure). Recently observed in the "PlasmaKristall-Experiment – Nefedov" experiments onboard the International Space Station and modeled analytically [14], this is the first treatment of the interaction by a self-consistent numerical model. The decrease in gelation time occurs due to the formation of high-density, high mass aggregates which act as seeds for runaway growth. This phenomenon is shown to occur even for a population of particles charged to a monopotential, which previously was assumed to retard coagulation rates.

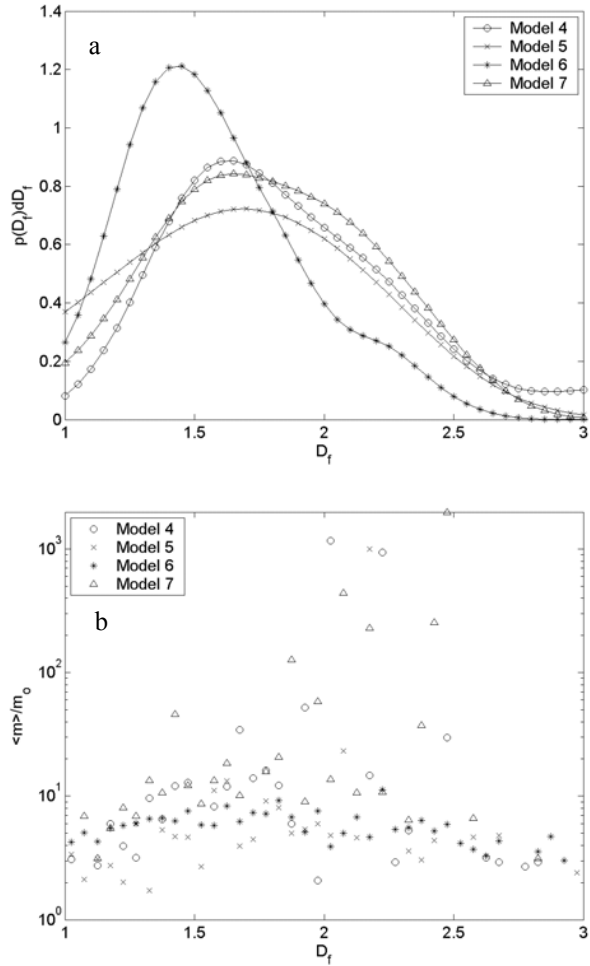

Fig. 7. Aggregate statistics for models 4-7, showing the effect of including the dipole moments of the aggregate and tree structure in the force calculations.

## REFERENCES

[1] S. V. W. Beckwith, T. Henning, and Y. Nakagawa, "Dust properties and assembly of large particles in protoplanetary disks," in *Protostars and Planets IV*, V. Mannings, A. P. Boss, and S. S. Russell, Eds., University of Arizona Press, pp. 533-558, 2000.

[2] L. Boufendi, and A. Bouchoule, "Particle nucleation and growth in a low-pressure argon-silane discharge," *Plasma Sources Sci. Technol.*, vol. 3, pp. 262-267, 1994.

[3] S. J. Weidenschilling, and J. N. Cuzzi, "Formation of planetesimals in the solar nebula.," in *Protostars and Planets III*, E. H. Levy and J. I. Lunine, Eds., University of Arizona Press, pp. 1031-1060, 1993.

[4] S. E. Strom, S. Edwards and M. F. Skrutskie, "Evolutionary time scales for circumstellar disks associated with intermediate- and solar-type stars," in *Protostars and Planets III*, E. H. Levy and J. I Lunine, Eds., University of Arizona Press, pp. 837-866, 1993.

[5] D. J. Hollenbach, H. W. Yorke, and D. Johnstone, "Disk dispersal around young stars," in *Protostars and Planets IV*, V. Mannings, A. P. Boss, and S. S. Russell, Eds., University of Arizona Press, pp. 401-428, 2000.

[6] D. J. Wilner and O. P. Lay, "Subarcsecond millimeter and submillimeter observations of circumstellar disk," in *Protostars and Planets IV*, V. Mannings, A. P. Boss, and S. S. Russell, Eds., University of Arizona Press, pp. 509-532, 2000.

[7] J. Blum and M. Münch, "Experimental investigations on aggregate-aggregate collisions in the early solar nebula," *Icarus*, vol. 106, pp. 151-167, 1993.

[8] J. Blum, G. Wurm, S. Kempf, and T. Henning, "The Brownian motion of dust particles in the solar nebula: an experimental approach to the problem of pre-planetary dust aggregation," *Icarus*, vol. 124, pp. 441-451, 1996.

[9] J. Blum, G. Wurm, T. Poppe, S. Kempf, and T. Kozasa, "First results from the cosmic dust aggregation experiment CODAG," *Adv. Space Res.*, vol. 29, no. 4, pp. 497-503, 2002.

[10] P. Meakin, and B. Donn, "Aerodynamic properties of fractal grains: implications for the primordial solar nebula," *Ap. J.*, vol. 329, pp. L39-L41, 1988.

[11] V. Ossenkopf, "Dust coagulation in dense molecular clouds: the formation of fluffy aggregates," *Astron. Astrophys.*, vol. 280, pp. 617-646, 1993.

[12] D. C. Richardson, "A self-consisten numerical treatment of fractal aggregate dynamics," *Icarus*, vol. 115, pp. 320-335, 1995.

[13] M. H. Lee, "On the validity of the coagulation equation and the nature of runaway growth," *Icarus*, vol. 143, pp. 74-86, 2000.

[14] A. V. Ivlev, G. E. Morfill and U. Konopka, "Coagulation of Charged Microparticles in Neutral Gas and Charge-Induced Gel Transitions," *Phys Rev Lett*, vol. 89, no. 19, 2002.

[15] L. Spitzer, *Physical Processes in the Interstellar Medium*, New York: John Wiley and Sons, 1978.

[16] A. Chokshi, A. G. G. M. Tielens, and D. Hollenbach, "Dust coagulation," *Ap. J.*, vol. 407, pp. 806-819, 1993.

[17] M. Horanyi and C. K. Goertz, "Coagulation of dust particles in a plasma," *Ap. J.*, vol. 361, pp. 155-161, 1990.

[18] V. W. Chow, D. A. Mendis, and M. Rosenburg, "Role of grain size and particle velocity distribution in secondary electron emission in space plasmas," *J. Geophys. Res.*, vol. 98, pp. 19,065-19,076, 1993.

[19] L. B. Barge and T. W. Hyde, "The Calculation of Grain Charge in a Dense Dusty Plasma With A Nonuniform Surface Potential," *Adv. Sp. Res.*, vol. 29, no. 9, pp. 1277-1282, 2002.

[20] L. B. Barge and T. W. Hyde, "Charging In A Dusty Plasma With A Size Distribution: A Comparison of Three Models," *Adv. Sp. Res.*, vol. 29, no. 9, pp. 1283-1288, 2002.

[21] L. B. Barge and T. W. Hyde, "A Charging Model For a Dust Cloud with a Size Distribution and A Nonuniform Potential," *Adv. Sp. Res.*, vol. 29, no. 9, pp. 1289-1294, 2002.

[22] I. A. Belov, A. S. Ivanov, D. A. Ivanov, et al., "Coagulation of Charged Particles in a Dusty Plasma," *J. Exp. And Theor. Phys.*, vol. 90, no. 1, pp. 93-101, 2000.

[23] S. Kempf, S. Pfalzner, and T. K. Henning, "N-Particle-Simulations of dust growth," *Icarus*, vol. 141, pp. 388-398, 1999.

[24] G. Wurm and J. Blum, "Experiments on preplanetary dust aggregation," *Icarus*, vol. 132, pp. 125-136, 1998.

[25] J. Blum, G. Wurm, S. Kempf, et al., "Growth and form of planetary seedlings: results from a microgravity aggregation experiment," *Phys Rev Lett*, vol. 85, no. 12, 2000.

[26] D. C. Richardson, "A new tree code method for simulation of planetesimal dynamics," *MNRAS*, vol. 261, pp. 396-414, 1993.

[27] D. C. Richardson, "Tree code simulations of planetary rings," *MNRAS*, vol. 269, pp. 493-511, 1994.

[28] J. Wisdom and S. Tremaine, "Local Simulations of Planetary Rings," *Astronom. J.*, vol. 95, no. 3, pp. 925-940, 1988.

[29] J. Barnes and P. Hut, "A hierarchical O(N log N) force-calculation algorithm," *Nature*, vol. 324, pp. 446-449, 1986.

[30] L. S. Matthews and T. W. Hyde, "Numerical modeling of gravito-electrodynamics in dusty plasmas," in *Proc. Of the Intl. Conf. on SCCS*, I. Kalman, ed., pp. 231-235, 1998.

[31] L. S. Matthews and T. W. Hyde, "Charged Grains In Saturn's F-Ring: Interaction With Saturn's Magnetic Field," *Adv. Sp. Res.*, in press, 2003.

[32] L. S. Matthews and T. W. Hyde, "Dynamics of Charged Grains in Saturn's F Ring: Encounters with Prometheus and Pandora," *J Phys A*, vol. 36, no. 22, pp. 6207-6214, 2003.

[33] L. B. Barge, L. S. Matthews and T. W. Hyde, "Coagulation in Dust Clouds Immersed in Transient Plasma Environments," *Adv. Sp. Res.*, in press, 2003.

[34] J. Vasut, T. W. Hyde and L. Barge, "Finite coulomb crystal formation," *Adv. Sp. Res.*, in press, 2003.

[35] J. Vasut and T. W. Hyde, "Computer simulations of coulmb crystallization in a dusty plasma," *IEEE Trans. Plasma Sci*, vol. 29, no. 2, pp. 231-237, 2001.

[36] K. Qiao and T. W. Hyde, "Numerical simulation and analysis of thermally excited waves in plasma crystals," *Adv. Sp. Res.*, in press, 2003.

[37] K. Qiao and T. W. Hyde, "Dispersion relations for thermally excited waves in plasma crystals," *J. Phys. A*, vol. 36, pp. 6109-6115, 2003.

[38] S. J. Weidenschilling, and T. V. Ruzmaikina, "Coagulation of grains in static and collapsing protostellar clouds," *Ap. J.*, vol. 430, pp. 713-726, 1994.

[39] S. J. Weidenschilling, "Evolution of grains in a turbulent solar nebula," *Icarus*, vol. 60, pp. 553-567, 1984.

[40] P. Meakin, "Fractal Aggregates in Physics," *Rev. Geophys*, vol. 29, pp. 317-354, 1991.

[41] B. Donn and J. M Duva, "Formation and properties of fluffy plantesimals," *Astrophys. and Sp. Sci.*, vol. 212, pp. 43-47, 1994.

[42] H. Kimura, H. Okamoto, and T. Mukai, "Radiation pressure and the Poynting-Robertson effect for fluffy dust particles," *Icarus*, vol. 157, pp. 349-361, 2002.

[43] P. Meakin, "Effects of cluster trajectories on cluster-cluster aggregation: A comparison of linear and Brownian trajectories in two- and three-dimensional simulations," *Phys Rev*, vol. A29, pp. 997-999, 1984.

[44] M. H. Ernst, in *Fractals in Physics*, Amsterdam: North Holland, pp. 289-302, 1986.

[45] B. B. Mandelbrot, *The Fractal Geometry of Nature*, San Francisco, CA: Freemont, 1982.